\pdfoutput=1

\documentclass[11pt]{article}

\usepackage[final]{acl}

\usepackage{times}
\usepackage{latexsym}

\usepackage[T1]{fontenc}

\usepackage[utf8]{inputenc}

\usepackage{microtype}

\usepackage{inconsolata}

\usepackage{graphicx}

\usepackage{amsfonts}
\usepackage{amsmath}
\usepackage{booktabs}
\usepackage{xspace}
\usepackage{algorithm}
\usepackage{comment}
\usepackage{adjustbox}
\usepackage{algpseudocode}
\usepackage{multirow}

\def\autoS{AutoScholarQuery\xspace}
\def\realS{RealScholarQuery\xspace}
\def\pasa{PaSa\xspace}

\title{PaSa: An LLM Agent for Comprehensive Academic Paper Search}

\author{
   \textbf{Yichen He$^*$}\textsuperscript{1}\quad
   \textbf{Guanhua Huang$^*$}\textsuperscript{1}  \quad
   \textbf{Peiyuan Feng}\textsuperscript{1} \quad
   \textbf{Yuan Lin$^\dagger$}\textsuperscript{1}\\[0.1cm]
   \textbf{Yuchen Zhang}\textsuperscript{1}\quad
   \textbf{Hang Li}\textsuperscript{1}\quad
   \textbf{Weinan E}\textsuperscript{2}\\[0.3cm]
   \textsuperscript{1}ByteDance Seed \quad \textsuperscript{2}Peking University\quad \\[0.3cm]
   \texttt{\{hyc,huangguanhua,fpy,linyuan.0\}@bytedance.com,}\\
   \texttt{\{zhangyuchen.zyc,lihang.lh\}@bytedance.com, weinan@math.pku.edu.cn}\\[0.3cm]
   Demo: \url{https://pasa-agent.ai}\\
}

\begin{document}
\maketitle

 \def\thefootnote{$*$}\footnotetext{Equal contribution.}

 \def\thefootnote{$\dagger$}\footnotetext{Corresponding author.}
\renewcommand{\thefootnote}{\arabic{footnote}}

\begin{abstract}
We introduce \pasa, an advanced {\bf Pa}per {\bf S}e{\bf a}rch agent powered by large language models. \pasa can autonomously make a series of decisions, including invoking search tools, reading papers, and selecting relevant references, to ultimately obtain comprehensive and accurate results for complex scholar queries. We optimize \pasa using reinforcement learning with a synthetic dataset, \autoS, which includes 35k fine-grained academic queries and corresponding papers sourced from top-tier AI conference publications. Additionally, we develop \realS, a benchmark collecting real-world academic queries to assess \pasa performance in more realistic scenarios. Despite being trained on synthetic data, \pasa significantly outperforms existing baselines on \realS, including Google, Google Scholar, Google with GPT-4o for paraphrased queries, ChatGPT (search-enabled GPT-4o), GPT-o1, and \pasa-GPT-4o (\pasa implemented by prompting GPT-4o). Notably, \pasa-7B surpasses the best Google-based baseline, Google with GPT-4o, by 37.78\% in recall@20 and 39.90\% in recall@50, and exceeds \pasa-GPT-4o by 30.36\% in recall and 4.25\% in precision. Model, datasets, and code are available at \url{https://github.com/bytedance/pasa}.

\end{abstract}

\section{Introduction}

Academic paper search lies at the core of research yet represents a particularly challenging information retrieval task. It requires long-tail specialized knowledge, comprehensive survey-level coverage, and the ability to address fine-grained queries. For instance, consider the query: \emph{"Which studies have focused on non-stationary reinforcement learning using value-based methods, specifically UCB-based algorithms?"} While widely used academic search systems like Google Scholar are effective for general queries, they often fall short when addressing these complex queries~\cite{gusenbauer2020academic}. Consequently, researchers frequently spend substantial time conducting literature surveys~\cite{kingsley2011not,gusenbauer2021every}.

The advancements in large language models (LLMs)~\cite{openai2023gpt4,anthropic2024claude,team2023gemini,yang2024qwen2} have inspired numerous studies leveraging LLMs to enhance information retrieval, particularly by refining or reformulating search queries to improve retrieval quality~\cite{alaofi2023can,li2023generate,ma2023query,peng2024large}. In academic search, however, the process goes beyond simple retrieval. Human researchers not only use search tools, but also engage in deeper activities, such as reading relevant papers and checking citations, to perform comprehensive and accurate literature surveys.

\begin{figure*}[!t] 
\centering 
\includegraphics[width=1.0\textwidth]{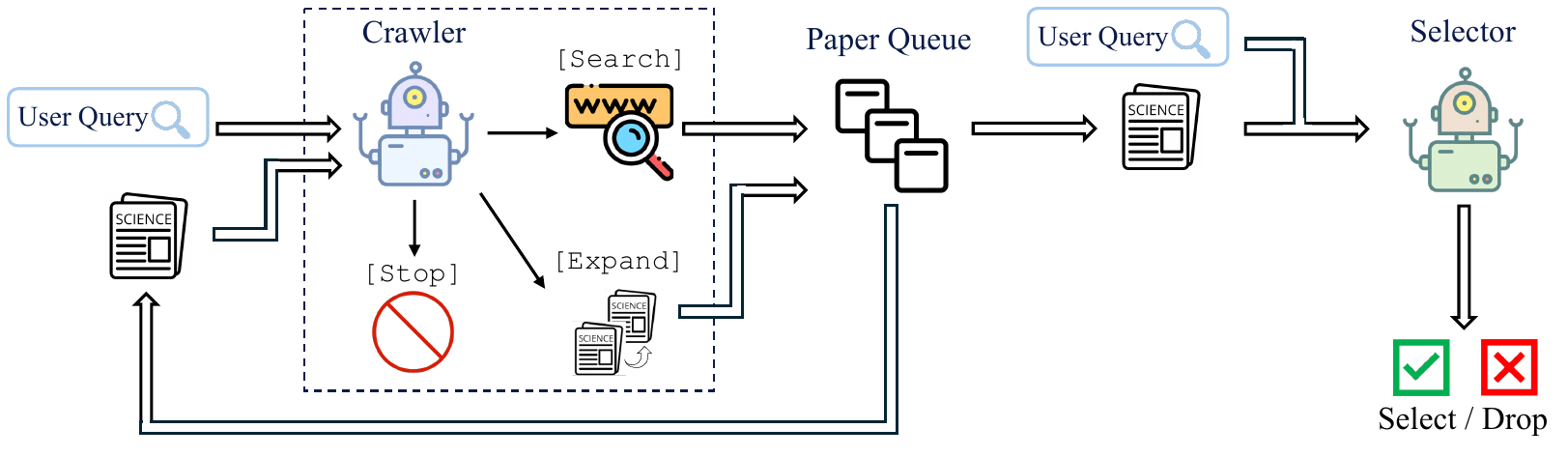} 
\caption{Architecture of \pasa. The system consists of two LLM agents, Crawler and Selector. The Crawler processes the user query and can access papers from the paper queue. It can autonomously invoke the search tool, expand citations, or stop processing of the current paper. All papers collected by the Crawler are appended to the paper queue. The Selector reads each paper in the paper queue to determine whether it meets the criteria specified in the user query.} 
\label{fig:motivation} 
\vspace{-10pt}
\end{figure*}

In this paper, we introduce \pasa, a novel paper search agent designed to mimic human behavior for comprehensive and accurate academic paper searches. As illustrated in Figure~\ref{fig:motivation}, PaSa consists of two LLM agents: the Crawler and the Selector. For a given user query, the Crawler can autonomously collect relevant papers by utilizing search tools or extracting citations from the current paper, which are then added to a growing \emph{paper queue}. The Crawler iteratively processes each paper in the paper queue, navigating citation networks to discover increasingly relevant papers. The Selector carefully reads each paper in the paper queue to determine whether it meets the requirements of the user query. We optimize \pasa within the AGILE, a reinforcement learning (RL) framework for LLM agents~\cite{feng2024agile}.

Effective training requires high-quality academic search data. Fortunately, human scientists have already created a vast amount of high-quality academic papers, which contain extensive surveys on a wide range of research topics. We build a synthetic but high-quality academic search dataset, \autoS, which collects fine-grained scholar queries and their corresponding relevant papers from the related work sections of papers published at ICLR 2023 \footnote{\url{https://iclr.cc/Conferences/2023}}, ICML 2023 \footnote{\url{https://icml.cc/Conferences/2023}}, NeurIPS 2023 \footnote{\url{https://neurips.cc/Conferences/2023}}, ACL 2024 \footnote{\url{https://2024.aclweb.org/}}, and CVPR 2024 \footnote{\url{https://cvpr.thecvf.com/Conferences/2024}}. \autoS includes 33,511 / 1,000 / 1,000 query-paper pairs in the training / development / test split. 

Although \autoS only provides query and paper answers, without demonstrating the path by which scientists collect the papers, we can utilize it to perform RL training to improve PaSa. In addition, we design a new session-level PPO (Proximal Policy Optimization~\cite{schulman2017proximal}) training method to address the unique challenges of the paper search task: 1) sparse reward: The papers in \autoS are collected via citations, making it a smaller subset of the actual qualified paper set. 2) long trajectories: The complete trajectory of the Crawler may involve hundreds of papers, which is too long to directly input into the LLM context.

To evaluate PaSa, besides the test set of \autoS, we also develop a benchmark, \realS. It contains 50 real-world academic queries with annotated relevant papers, to assess \pasa in real-world scenarios. We compare \pasa with several baselines including Google, Google Scholar, Google paired with GPT-4o for paraphrased queries, ChatGPT (search-enabled GPT-4o), GPT-o1 and \pasa-GPT-4o (\pasa agent realized by prompting GPT-4o). Our experiments show that \pasa-7b significantly outperforms all baselines. Specifically, for \autoS test set, \pasa-7b achieves a 34.05\% improvement in Recall@20 and a 39.36\% improvement in Recall@50 compared to Google with GPT-4o, the strongest Google-based baseline. \pasa-7b surpasses \pasa-GPT-4o by 11.12\% in recall, with similar precision. For \realS, \pasa-7b outperforms Google with GPT-4o by 37.78\% in Recall@20 and 39.90\% in Recall@50. \pasa-7b surpasses \pasa-GPT-4o by 30.36\% in recall and 4.25\% in precision.

The main contributions of this paper are summarized as follows:
\begin{itemize}
    \item We introduce \pasa, a comprehensive and accurate paper search agent that can autonomously use online search tools, read entire papers, and navigate citation networks.
    \item We develop two high-quality datasets for complex academic search, \autoS and \realS.
    \item Although \pasa is trained solely on synthetic data, it achieves remarkable real-world performance. Experiments demonstrate that \pasa, built on 7B LLM, significantly outperforms all baselines, including GPT-4 agent, Google-based search, and ChatGPT.

\end{itemize}

\section{Related Work}

\paragraph{LLMs in Scientific Discovery}

LLMs have been applied across various stages of scientific discovery~\cite{van2023ai,lu2024ai,messeri2024artificial,liao2024llms}, such as brainstorming ideas~\cite{girotra2023ideas,wang-etal-2024-scimon,baek2024researchagent}, designing experiments~\cite{m2024augmenting}, writing code~\cite{xu2022systematic}, and generating research papers~\cite{shao-etal-2024-assisting,agarwal2024litllm,wang2024autosurvey}. One of the most fundamental yet critical stages in research is conducting academic surveys. Despite its importance, current tools like Google Scholar are often insufficient, leading researchers to spend considerable time on literature review tasks~\cite{kingsley2011not,gusenbauer2021every,gusenbauer2020academic}. This challenge motivates us to develop PaSa, an LLM agent designed to autonomously and comprehensively assist researchers in collecting relevant research papers for complex scholarly queries.

\paragraph{LLM Agents}
LLM Agents combine LLMs with memory, tool use, and planning, enabling them to perform more complex tasks such as personal copilots~\citep{stratton2024introduction}, travel planning~\citep{gundawar2024robust}, web operations~\citep{deng2024mind2web}, software development~\citep{qian2023communicative}, and scientific experimentation~\citep{bran2023chemcrow}. In addition to realizing LLM Agents through prompt engineering~\cite{park2023generative,yao2023react,shinn2024reflexion,chen2023autoagents}, recent research has focused on optimizing and training these agents~\cite{feng2024agile,putta2024agent,liu2023reason}. Among these efforts, AGILE~\cite{feng2024agile}, a reinforcement learning framework for LLM agents, allows the joint optimization of all agent skills in an end-to-end manner. In our work, we adopt the AGILE framework to implement PaSa. Specifically, we design a novel session-level PPO algorithm to address the unique challenges of the paper search task, including sparse rewards and long trajectories.

\section{Datasets}

\subsection{\autoS} \label{sec:AutoScholarQuery}

\autoS is a synthetic but high-quality dataset of academic queries and related papers, specifically curated for the AI field.

To construct \autoS, we began by collecting all papers published at ICLR 2023, ICML 2023, NeurIPS 2023, ACL 2024, and CVPR 2024. For the Related Work section of each paper, we prompted GPT-4o~\cite{hurst2024gpt} to generate scholarly queries, where the answers to these queries correspond to the references cited in the Related Work section. The prompt used is shown in Appendix~\ref{autoS_prompt}. For each query, we retained only the papers that could be retrieved on arXiv\footnote{\url{https://arxiv.org/}}, using their \texttt{arxiv\_id} as the unique article identifier in the dataset. We adopt the publication date of the source paper as the query date. During both training and testing, we only considered papers published prior to the query date.

The final \autoS dataset comprises 33,551, 1,000, and 1,000 instances in the training, development, and testing splits, respectively. Each instance consists of a query, the associated paper set, and the query date, with queries in each split derived from distinct source papers. Table~\ref{tab:autoS_examples} provides illustrative examples from \autoS, while additional dataset statistics are summarized in Table~\ref{tab:dataset-statistics}.

To evaluate the quality of \autoS, we sampled 100 query-paper pairs and assessed the rationality and relevance of each query and the corresponding paper. A qualified query should be meaningful and unambiguous. A qualified paper should match the requirements of the scholarly query. Detailed evaluation criteria are provided in Appendix.~\ref{auto_eval}. Three authors manually reviewed each pair, determining that 94.0\% of the queries were qualified. Among these qualified queries, 93.7\% had corresponding papers that were deemed relevant and appropriate.

\begin{table*}[!h]
\centering
\scalebox{0.75}{
\begin{tabular}{|p{1.25\textwidth}|}
\hline
\begin{tabular}[c]{@{}p{1.25\textwidth}@{}}  \textbf{Query:} Could you provide me some studies that proposed hierarchical neural models to capture spatiotemporal features in sign videos?\\
\textbf{Query Date:} 2023-05-02 \\
\textbf{Answer Papers:}
\\ {[}1{]} TSPNet: Hierarchical Feature Learning via Temporal Semantic Pyramid for Sign Language Translation (2010.05468)
\\ {[}2{]} Sign Language Translation with Hierarchical Spatio-Temporal Graph Neural Network (2111.07258)
\end{tabular} 
\\ 
\textbf{Source:} SLTUnet: A Simple Unified Model for Sign Language Translation, ICLR 2023\\

\hline
\begin{tabular}[c]{@{}p{1.25\textwidth}@{}}   \textbf{Query:} Which studies have focused on nonstationary RL using value-based methods, specifically Upper Confidence Bound (UCB) based algorithms?\\ 
\textbf{Query Date:} 2023-08-10 \\
\textbf{Answer Papers:}
\\ {[}1{]} Reinforcement Learning for Non-Stationary Markov Decision Processes: The Blessing of (More) Optimism (2006.14389)
\\ {[}2{]} Efficient Learning in Non-Stationary Linear Markov Decision Processes (2010.12870)
\\ {[}3{]} Nonstationary Reinforcement Learning with Linear Function Approximation (2010.04244) \\
\textbf{Source:} Provably Efficient Algorithm for Nonstationary Low-Rank MDPs, NeurIPS 2023\\ \end{tabular} \\ 
\hline
\begin{tabular}[c]{@{}p{1.25\textwidth}@{}}
\textbf{Query:} Which studies have been conducted in long-form text generation, specifically in story generation?\\
\textbf{Query Date:} 2024-01-26 \\
\textbf{Answer Papers:}
\\ {[}1{]} Strategies for Structuring Story Generation (1902.01109)
\\ {[}2{]} MEGATRON-CNTRL: Controllable Story Generation with External Knowledge Using Large-Scale Language Models (2010.00840) \\
\textbf{Source:} ProxyQA: An Alternative Framework for Evaluating Long-Form Text Generation with Large Language Models, ACL 2024\\
\end{tabular} \\ \hline
\end{tabular}
}
\caption{Examples of queries and corresponding papers in \autoS.}
\label{tab:autoS_examples}
\end{table*}

\begin{table}[!h]
\centering
\scalebox{0.72}{
\begin{tabular}{lccccc}
\toprule
Conference & $|P|$ & $|Q|$ & $Ans(/Q)$ & $Ans$-$50$ & $Ans$-$90$ \\ \midrule
ICLR 2023    & 888  & 5204 & 2.46 & 2.0 & 5.0 \\
ICML 2023    & 981  & 5743 & 2.37 & 2.0 & 5.0 \\
NeurIPS 2023 & 1948 & 11761 & 2.59 & 2.0 & 5.0 \\
CVPR 2024    & 1336 & 9528 & 2.94 & 2.0 & 6.0 \\
ACL 2024     & 485  & 3315 & 2.16 & 2.0 & 4.0 \\
\bottomrule
\end{tabular}
}
\caption{Statistics of \autoS. $|P|$ and $|Q|$ represent the total number of papers and queries collected for each conference. $Ans(/Q)$ denotes the average number of answer papers per query. $Ans$-$50$ and $Ans$-$90$ refers to the 50th and 90th percentiles of answer paper counts per query.}
\label{tab:dataset-statistics}
\end{table}

\begin{figure*}[h]
    \centering
    \includegraphics[width=1.0\linewidth]{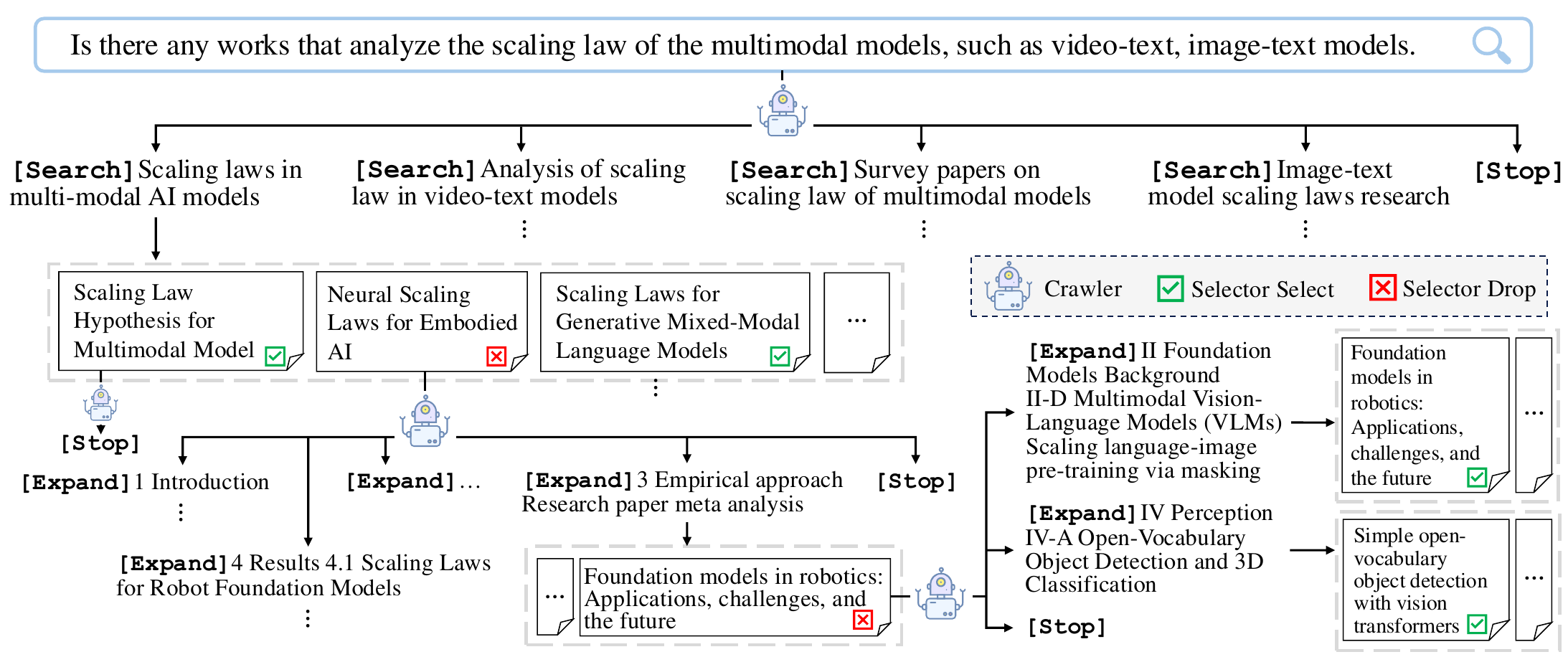}
    \caption{An example of the \pasa workflow. The Crawler runs multiple \texttt{[Search]} using diverse and complementary queries. In addition, the Crawler can evaluate the long-term value of its actions. Notably, it discovers many relevant papers as it explores deeper in the citation network, even when intermediate papers along the path do not align with the user query.}
    \label{search_tree}
\vspace{-10pt}
\end{figure*}

\subsection{\realS}

To evaluate \pasa in more realistic scenarios, we constructed \realS, a test dataset consisting of 50 real-world research queries. After launching the demo of \pasa, we invited several AI researchers to use the system. From the queries they provided, we randomly sampled a subset of queries and manually filtered out overly broad topics (e.g., "multimodal large language models," "video generation"). Ultimately, we collected 50 fine-grained and realistic queries.

For each query, we first manually gathered relevant papers to the best of our ability. To ensure comprehensive coverage, we then applied multiple methods to retrieve additional papers, including \pasa, Google, Google Scholar, ChatGPT (search-enabled GPT-4o), and Google paired with GPT-4o for paraphrased queries. As these methods also serve as baselines for comparison with \pasa, implementation details are deferred to Section~\ref{baselines}. The results from all methods were aggregated into a pool of candidate papers. Finally, professional annotators reviewed all candidate papers for each query, selecting those that met the specific requirements of the query to create the final set of relevant papers. Annotation guidelines and quality control procedures are detailed in Appendix.~\ref{annotation_details}. The query date of all instances in \realS is 2024-10-01. Table~\ref{tab:realS_examples} in Appendix~\ref{data_example} provides an example from \realS.

The annotators included professors from the Department of Computer Science at a top-tier university in China. On average, each query required the annotators to review 76 candidate papers. We paid \$4 per data entry (a query-paper pair), resulting in an average of \$304 per query. Given the high annotation cost, we completed this process for only 50 instances. On average, each query is associated with 15.82 answer papers. The 50th percentile of answer counts per query is 9, while the 90th percentile reaches 37.

\section{Methodology}

\subsection{Overview}

As illustrated in Figure~\ref{fig:motivation}, the \pasa system consists of two LLM agents: Crawler and Selector. The crawler reads the user's query, generates multiple search queries, and retrieves relevant papers. The retrieved papers are added to a \emph{paper queue}. The Crawler further processes each paper in the paper queue to identify key citations worth exploring further, appending any newly relevant papers to the paper queue. The selector conducts a thorough review of each paper in the paper queue to assess whether it fulfills the user's query requirements.

In summary, the Crawler is designed to maximize the recall of relevant papers, whereas the Selector emphasizes precision in identifying papers that meet the user's needs.

\begin{table}[ht]
\renewcommand{\arraystretch}{1}
\centering
\resizebox{0.48\textwidth}{!}{
\begin{tabular}{ll}
\toprule[1pt]
\multicolumn{1}{c}{\textbf{Name}} & \multicolumn{1}{c}{\textbf{Implementation}} \\
\midrule[0.5pt]
& Generate a search query and invoke \\ 
\texttt{[Search]} & the search tool. Append all resulting \\ 
& papers to the paper queue. \\
\midrule[0.5pt]
& Generate a subsection name, then \\ 
\texttt{[Expand]} & add all referenced papers in the sub- \\ 
& section to the paper queue.\\
\midrule[0.5pt]
\multirow{2}{*}{\texttt{[Stop]}} & Reset the context to the user query and \\ 
& the next paper in the paper queue. \\ 
\bottomrule[1pt]
\end{tabular}
}
\caption{Functions of the Crawler.}
\label{agent_function_table}
\vspace{-15pt}
\end{table}

\subsection{Crawler}

In RL terminology, the Crawler performs a token-level Markov Decision Process (MDP). The action space $\mathcal{A}$ corresponds to the LLM's vocabulary, where each token represents an action. The LLM functions as the policy model. The agent's state is defined by the current LLM context and the paper queue. The Crawler operates with three registered functions, as outlined in Table~\ref{agent_function_table}. When an action matches a function name, the corresponding function is executed, further modifying the agent's state.

For example, as Figure~\ref{search_tree} shows, the agent begins by receiving a user query, incorporating it into its context, and initiating actions. If the token generated is \texttt{[Search]}, the LLM continues to generate a search query, and the agent invokes a search tool to retrieve papers, which are then added to the paper queue. If the token is \texttt{[Expand]}, the LLM continues to extract a subsection name from the current paper in its context. The agent then extracts all referenced papers within that subsection, adding them to the paper queue. If the token is \texttt{[Stop]}, the agent resets its context to the user query and information of the next paper in the paper queue. This information includes the title, abstract, and an outline of all sections and subsections.

The training process for the Crawler comprises two stages. In the first stage, we generate trajectories for a small subset of the training data and then perform imitation learning (see Appendix~\ref{crawler_detail} for details).  In the second stage, reinforcement learning is applied. The details of the RL training implementation are described below.

\paragraph{Reward Design}

We conduct RL training on the \autoS training set, where each instance consists of a query $q$ and a corresponding paper set $\mathcal{P}$. 
Starting with a query $q$, the Crawler generates a trajectory $\tau=(s_1, a_1, \cdots, s_T, a_T)$. At each time step $t$, we denote the current paper queue as $\mathcal{Q}_t$. Upon taking action $a_t$, the Crawler appends a set of new papers $(p_1, p_2, \cdots, p_{n_t})$ to the paper queue. If $a_t = \texttt{[Stop]}$, the set is empty and no papers are added.

The reward of executing action $a_t$ in state $s_t$ is defined as

\begin{small}
\begin{equation}\label{action_reward}
    r(s_t, a_t) = \alpha\times\sum_{i=1}^{n_t} \mathbb{I}(q, p_i, t) - c(a_t),
\end{equation}
\end{small}
where $\mathbb{I}(q, p_i, t)=1$ if $p_i$ matches the query $q$ and is not already in $\mathcal{Q}_t$, and $\mathbb{I}(q, p_i, t)=0$ otherwise. Here, $\alpha$ is a reward coefficient, and $c(a_t)$ is the cost of action $a_t$.

The indicator function $\mathbb{I}(q, p_i, t)$ can be determined by checking if $p_i$ belongs to $\mathcal{P}-\mathcal{Q}_t$. However, it is important to note that the \autoS may only include a subset of the ground-truth papers, as citations often emphasize a limited number of key references. If the Crawler receives rewards solely based on matching papers in \autoS, this could lead to sparse rewards during training. To mitigate this, we use the Selector as an auxiliary reward model for the Crawler. The revised definition of $\mathbb{I}(q, p_i, t)$ is:

\begin{small}
\begin{equation}
\mathbb{I}(q, p_i, t) = 
\begin{cases}
1, & \text{if } \left( \text{Selector}(q, p_i) = 1 \text{ or } p_i \in \mathcal{P} \right) \\
& \quad \text{and } p_i \notin \mathcal{Q}_t, \\
0, & \text{otherwise.}
\end{cases}
\end{equation}
\end{small}
Here $\text{Selector}(q, p_i)=1$ if paper $p_i$ is identified as correct to meet the query $q$ by the Selector, and $\text{Selector}(q, p_i)=0$ otherwise.

\paragraph{RL Training}

A key challenge in training the Crawler with RL is the significant time required to sample a complete trajectory for a given query. This is due to each \texttt{[Search]} or \texttt{[Expand]} action adding multiple papers to the paper queue, resulting in hundreds or even thousands of papers in the final paper queue. 



To address this issue, we define a \emph{session} as a sub-trajectory that ends with the \texttt{[Stop]} action, after which a new session begins. We identify two types of initial states for such sub-trajectories: $S_q$, containing only the user query, and $S_{q+p}$, containing both the query and a paper. $S_q$ represents the task's starting point, where the LLM context includes only the query. In contrast, $S_{q+p}$ arises after a \texttt{[Stop]} action, where the LLM context is reset to the query and the next paper in the queue.

Formally, we model the Crawler as a policy $\pi_\theta(a_t|s_t)$. We partition the entire trajectory $\tau=(s_1, a_1, \cdots, s_T, a_T)$ into a sequence of sessions: $(\tau_{t_1:t_2-1},\tau_{t_2:t_3-1},\cdots)$. Each session is $\tau_{t_i:t_{i+1}-1}=(s_{t_i}, a_{t_i}, \cdots, s_{t_{i+1}-1}, a_{t_{i+1}-1})$, where the initial state $s_{t_i}$ is either belonging to type $S_q$ or $S_{q+p}$, and the final action $a_{t_{i+1}-1}$ is \texttt{[STOP]}.

Sampling such a sub-trajectory from these session initial states is computationally efficient. During the PPO training, at time step $t\in[t_i, t_{i+1})$, we estimate the return in the session using Monte Carlo sampling:

\begin{small}
\begin{eqnarray}\label{q_function}
\hat{R}_t &=& \sum_{k=t}^{t_{i+1}-1}\gamma_0^{k-t}\bigg[r(s_{k}, a_{k}) \\
&& +\gamma_1\sum_{j=1}^{n_{k}}\hat{V}_\phi(S_{q+p_j})\bigg] - \beta\cdot\log\frac{\pi_\theta(a_t|s_t)}{\pi_{\rm sft}(a_t|s_t)}\nonumber
\end{eqnarray}
\end{small}
Here, $\gamma_0$ is the in-session discount factor, and $\gamma_1$ is the across-session discount factor. $\hat{V}_\phi(\cdot)$ is the value function model to approximate the state value. After executing $a_{k}$, the paper queue is updated to include the newly found papers $(p_1, p_2, \cdots, p_{n_{k}})$. Since the Crawler will subsequently initiate new sessions to process these additional papers, their associated reward-to-go should be incorporated into the return estimate. In addition, we include a per-token KL penalty term from the learned policy $\pi_\theta$ to the initial policy $\pi_{\rm sft}$ obtained through imitation learning at each token to mitigate over-optimization. This term is scaled by the coefficient $\beta$.

Then the advantage function can be approximated by

\begin{small}
\begin{eqnarray}\label{advantage}
\hat{A}(s_t, a_t)&=& \hat{R}_t-\hat{V}_\phi(s_t).
\end{eqnarray}
\end{small}

Finally, the policy and value objectives can be given by

\begin{small}
\begin{eqnarray}\label{policy_loss}
\mathcal{L}_{\text{policy}}(\theta)=\!\mathbb{E}_{\tau'\sim\pi_{\theta}^\text{old}}\Bigg[\min\bigg(\frac{\pi_\theta(a_t|s_t)}{\pi_{\theta}^\text{old}(a_t|s_t)}\hat{A}(s_t, a_t),\\
\text{clip}\Big(\frac{\pi_\theta(a_t|s_t)}{\pi_{\theta}^\text{old}(a_t|s_t)}, 1-\epsilon, 1+\epsilon\Big)\hat{A}(s_t, a_t)\bigg)\Bigg]\nonumber
\end{eqnarray}
\end{small}
and

\begin{small}
\begin{eqnarray}\label{value_loss}
\mathcal{L}_{\text{value}}(\phi)=&\mathbb{E}_{\tau'\sim\pi_{\theta}^\text{old}}\Bigg[\rm{max}\bigg(\Big(\hat{R}_t-\hat{V}_\phi(s_t)\Big)^2,\\\nonumber
&\Big(\hat{R}_t-\hat{V}_{\phi}^{\rm{clip}}(s_t)\Big)^2\bigg)\Bigg],
\end{eqnarray}
\end{small}
respectively, where
\begin{small}
\begin{eqnarray}
    \hat{V}_{\phi}^{\rm{clip}}(s_t)=&\text{clip}\Big(\hat{V}_{\phi}(s_t),V_{\phi}^{\text{old}}(s_t)-\epsilon,V_{\phi}^{\text{old}}(s_t)+\epsilon\Big).
\end{eqnarray}
\end{small}
Here, $\pi_{\theta}^\text{old}$ and $V_{\phi}^{\text{old}}$ is used for sampling and $\tau'$ is session trajectory. We then combine these into the unified RL loss:

\begin{small}
\begin{eqnarray}\label{rl_loss}
\mathcal{L}_{\text{RL}}(\theta, \phi)=\mathcal{L}_{\text{policy}}(\theta)+\eta\cdot\mathcal{L}_{\text{value}}(\phi)
\end{eqnarray}
\end{small}
where $\eta$ is the coefficient of the value objective. 

\subsection{Selector}

The Selector is an LLM agent that takes two inputs: a scholar query and a research paper (including its title and abstract). It generates two outputs: (1) a single decision token $d$, either "True" or "False", indicating whether the paper satisfies the query, and (2) a rationale $r=(r_1, r_2, ..., r_{m})$ containing $m$ tokens that support this decision. The rationale serves two purposes: enhancing decision accuracy by jointly training the model to generate decisions and explanations, and improving user trust by providing the reasoning in \pasa application. 

To optimize training efficiency for the Crawler, the decision token is presented before the rationale, allowing the Selector to act as a single-token reward model during the Crawler training. Additionally, the token probability of the decision token can be used to rank search results. At last, as shown in Table~\ref{tab:selector_evaluation}, the order of the decision and rationale does not affect the Selector's performance.

We perform imitation learning to optimize the Selector. See Appendix~\ref{selector_detail} for training data collection and training details.

\section{Experiments}

\subsection{Experimental Setting}\label{experimental_setting}

We sequentially trained the Selector and Crawler, both based on the Qwen2.5-7b~\cite{yang2024qwen2}, to develop the final agent, referred to as \pasa-7b. 

\paragraph{Selector}

The Selector was fine-tuned using the training dataset described in Appendix~\ref{selector_detail}. We conducted supervised fine-tuning for one epoch with a learning rate of 1e-5 and a batch size of 4. The training runs on 8 NVIDIA-H100 GPUs.

\paragraph{Crawler}
The training process involves two stages. First, we perform imitation learning for 1 epoch on 12,989 training data with a learning rate of 1e-5 and batch size of 4 per device, using 8 NVIDIA H100 GPUs. In the second stage, we apply PPO training. To ensure stability, we first freeze the policy model and train the value model, followed by co-training both the policy and value models. The hyperparameters used during the training process are listed in Table~\ref{tab:crawler_hyperparameters} in Appendix~\ref{ppo_training_detail}. 

During imitation learning, the model encounters 5,000 queries, while during the RL training phase, the model processes a total of 16,000 queries. For more details please refer to Appendix~\ref{crawler_detail} for the imitation learning data construction and Appendix~\ref{ppo_training_detail} for the PPO training data sampling.

\paragraph{Implementation of \texttt{[Search]}}
The LLM predicts a query based on the context. Then the agent calls Google\footnote{Accessed via the Google Search API provided by \url{https://serper.dev}.\label{serper_api}} with the parameters \texttt{site:arxiv.org} and \texttt{before:query\_date}, restricting search results by source and publication time. 

\paragraph{Paper Management} 
We developed a database to manage and restore research papers. \pasa retrieves paper information from the database. If no matching record is found, we use ar5iv\footnote{\url{https://ar5iv.org/}} to obtain the full paper content, including citations, and then parse this data and store it in the database.

\begin{table*}[htbp]
\renewcommand{\arraystretch}{1}
\centering
\scalebox{0.85}{
\begin{tabular}{lcccccc}
\toprule[1pt]
\multicolumn{1}{l}{\textbf{Method}} & \textbf{Crawler Recall} & \textbf{Precision} & \textbf{Recall} & \textbf{Recall@100} & \textbf{Recall@50} & \textbf{Recall@20} \\
\midrule[0.5pt]
Google & - & - & - & 0.2015 & 0.1891 & 0.1568\\
Google Scholar & - & - & - & 0.1130 & 0.0970 & 0.0609\\
Google with GPT-4o & - & - & - & 0.2683 & 0.2450 & 0.1921\\
ChatGPT* & - & 0.0507 & 0.3046 & - & - & - \\
GPT-o1 & - & 0.0413 & 0.1925 & - & - & - \\
\pasa-GPT-4o & 0.7565 & \textbf{0.1457} & 0.3873 & - & - & - \\
\midrule[0.5pt]
\pasa-7b & \textbf{0.7931} & \textbf{0.1448} & \textbf{0.4834} & \textbf{0.6947} & \textbf{0.6334} & \textbf{0.5301} \\
\pasa-7b-ensemble & \textbf{0.8265} & 0.1410 & \textbf{0.4985} & \textbf{0.7099} & \textbf{0.6386} & \textbf{0.5326} \\
\bottomrule[1pt]
\end{tabular}
}
\caption{Results on AutoScholarQuery test set. *: Due
to the need for manual query submission, the ChatGPT baseline is evaluated on 100 randomly sampled instances. Results for all
methods on this subset are reported in Table~\ref{main_results_100}.}
\label{main_results}
\vspace{-0.2cm}
\end{table*}

\begin{table*}[htbp]
\renewcommand{\arraystretch}{1}
\centering
\scalebox{0.85}{
\begin{tabular}{lcccccc}
\toprule[1pt]
\multicolumn{1}{l}{\textbf{Method}} & \textbf{Crawler Recall} & \textbf{Precision} & \textbf{Recall} & \textbf{Recall@100} & \textbf{Recall@50} & \textbf{Recall@20} \\
\midrule[0.5pt]
Google & - & - & - & 0.2535 & 0.2342 & 0.1834\\
Google Scholar & - & - & - & 0.2809 & 0.2155 & 0.1514\\
Google with GPT-4o & - & - & - & 0.2946 & 0.2573 & 0.2020\\
ChatGPT & - & 0.2280 & 0.2007 & - & - & - \\
GPT-o1 & - & 0.058 & 0.0134 & - & - & - \\
\pasa-GPT-4o & 0.5494 & 0.4721 & 0.3075 & - & - & - \\
\midrule[0.5pt]
\pasa-7b & \textbf{0.7071} & \textbf{0.5146} & \textbf{0.6111} & \textbf{0.6929} & \textbf{0.6563} & \textbf{0.5798} \\
\pasa-7b-ensemble & \textbf{0.7503} & \textbf{0.4938} & \textbf{0.6488} & \textbf{0.7281} & \textbf{0.6877} & \textbf{0.5986} \\
\bottomrule[1pt]
\end{tabular}
}
\caption{Results on RealScholarQuery.}
\label{main_results_real}
\vspace{-0.2cm}
\end{table*}

\subsection{Baselines and Evaluation}\label{baselines}

We evaluate our paper search agent on both the test set of \autoS and \realS. We compare \pasa-7b against the following baselines:

\begin{itemize}

\item{\bf Google.} We use Google to search the query directly, with the same parameter settings in Section~\ref{experimental_setting}.

\item{\bf Google Scholar.} Queries are submitted directly to Google Scholar\textsuperscript{\ref{serper_api}}, with the same parameter settings in Section~\ref{experimental_setting}.

\item{\bf Google with GPT-4o.} We first employ GPT-4o to paraphrase the scholar query. The paraphrased query is then searched on Google.

\item{\bf ChatGPT.} We submit scholar query to ChatGPT\footnote{\url{https://chatgpt.com}}, powered by search-enabled GPT-4o.

\item{\bf GPT-o1.} Prompt GPT-o1 to process the scholar query. Note that it does not have access to external search tools.

\item{\bf \pasa-GPT-4o.} Implement \pasa as illustrated in Figure~\ref{fig:motivation} by prompting GPT-4o. It can perform multiple searches, paper reading, and citation network crawling. 
\end{itemize}

We carefully designed prompts for all baselines and they are shown in Appendix~\ref{prompt_templates}. All baselines, except \pasa-GPT-4o, represent the best-known scholar search methods. These comparisons highlight the effectiveness of our agentic approach. The comparison with \pasa-GPT-4o isolates the impact of RL training. 


As shown in Figure~\ref{search_tree}, the crawling process of \pasa can be visualized as a paper tree. In practice, considering the computational expense, we limit the Crawler's exploration depth to three for both \pasa-7b and \pasa-GPT-4o. 

For Google-based baselines, we evaluate recall using Recall@20, Recall@50, and Recall@100 metrics for the top-20, top-50, and top-100 search results, respectively. For other baselines that do not produce rankings, we assess precision and recall for the final retrieved papers. Additionally, we compare Crawler recall between \pasa-GPT-4o and \pasa-7b, defined as the proportion of target papers collected by the Crawler. This measures how many target papers are successfully included in the paper queue generated by the Crawler.


\begin{table}[htbp]
\renewcommand{\arraystretch}{1}
\centering
\scalebox{0.75}{
\begin{tabular}{lccc}
\toprule[1pt]
\multicolumn{1}{l}{\textbf{Method}} & \textbf{Precision} & \textbf{Recall} & \textbf{F1} \\
\midrule[0.5pt]
GPT-4o & 0.96 & 0.69 & 0.80 \\
Qwen-2.5-7b & 1.0 & 0.38 & 0.55 \\
\pasa-7b-Selector & 0.95 & 0.78 & \textbf{0.85} \\
\pasa-7b-Selector (Reason First) & 0.94 & 0.76 & \textbf{0.84} \\
\bottomrule[1pt]
\end{tabular}
}
\caption{Selector Evaluation.}
\label{tab:selector_evaluation}
\vspace{-15pt}
\end{table}

\subsection{Main results}

\begin{table*}[htbp]
\renewcommand{\arraystretch}{1}
\centering

\scalebox{0.85}{
\begin{tabular}{lcccccc}
\toprule[1pt]
\multirow{2}{*}{\textbf{Method}} & \multicolumn{3}{c}{\textbf{AutoScholarQuery}} & \multicolumn{3}{c}{\textbf{RealScholarQuery}}  \\ \cline{2-7} & \textbf{Crawler Recall} & \textbf{Precision} & \textbf{Recall} & \textbf{Crawler Recall} & \textbf{Precision} & \textbf{Recall} \\
\midrule[0.5pt]
w/o \texttt{[Expand]} & 0.3355 & 0.1445 & 0.2536 & 0.3359 & \textbf{0.6738} & 0.2890 \\
w/o RL training & 0.6556 & 0.1476 & 0.4210 & 0.4847 & 0.5155 & 0.4115 \\
w/o Selector as RM & 0.7041 & \textbf{0.1535} & 0.4458 & 0.5994 & 0.5489 & 0.5148 \\
\midrule[0.5pt]
\pasa-7b & \textbf{0.7931} & 0.1448 & \textbf{0.4834} & \textbf{0.7071} & 0.5146 & \textbf{0.6111} \\
\bottomrule[1pt]
\end{tabular}
}
\caption{Ablation study results on \autoS test set and \realS.}
\label{ablation}
\vspace{-0.2cm}
\end{table*}


As shown in Table~\ref{main_results}, \pasa-7b outperforms all baselines on \autoS test set. Specifically, compared to the strongest baseline, \pasa-GPT-4o, \pasa-7b demonstrates a 9.64\% improvement in recall with comparable precision. Moreover, the recall of the Crawler in \pasa-7b is 3.66\% higher than that in \pasa-GPT-4o. When compared to the best Google-based baseline, Google with GPT-4o, \pasa-7b achieves an improvement of 33.80\%, 38.83\% and 42.64\% in Recall@20, Recall@50 and Recall@100, respectively.

We observe that using multiple ensembles of Crawler during inference can improve performance. Specifically, we use sampling decoding to run Crawler twice in the \pasa-7b-ensemble setting, which boosts Crawler recall by 3.34\% on \autoS and increases the final recall by 1.51\%, with no significant change in precision. 

To evaluate \pasa in a more realistic setting, we assess its effectiveness on \realS. As illustrated in Table~\ref{main_results_real}, \pasa-7b exhibits a greater advantage in real-world academic search scenarios. Compared to \pasa-GPT-4o, \pasa-7b achieves improvements of 30.36\% in recall and 4.25\% in precision. Against the best Google-based baseline on \realS, Google with GPT-4o, \pasa-7b outperforms Google by 37.78\%, 39.90\%, and 39.83\% in recall@20, recall@50 and recall@100, respectively. Additionally, the \pasa-7b-ensemble further enhances crawler recall by 4.32\%, contributing to an overall 3.52\% improvement in the recall of the entire agent system.

As both the final decision-maker and auxiliary reward model in RL training for the Crawler, the performance of the Selector is crucial. To evaluate its effectiveness, we collected a dataset of 200 query-paper pairs, annotating whether each paper meets the query's requirements. This dataset serves as the benchmark for evaluating the Selector (see Appendix~\ref{sec:Details of Selector Evaluation} for details). We then compared our Selector against GPT-4o~\cite{hurst2024gpt} and Qwen-2.5-7b~\cite{yang2024qwen2}, as shown in Table~\ref{tab:selector_evaluation}. The results show that our Selector achieves an F1 score of 85\%, outperforming GPT-4o by 5\% and Qwen-2.5-7b by 30\%. Additionally, when compared to a setting where reasoning precedes decision token generation, the performance is comparable. Lastly, the Selector’s precision reaches 95\%, confirming its effectiveness as an auxiliary reward model for the Crawler RL training.

\subsection{Ablation study}

We perform ablation studies in Table~\ref{ablation} to evaluate the individual contributions of exploring citation networks, RL training, and using the Selector as the reward model. The results indicate that removing the \texttt{[Expand]} action from the Crawler leads to a significant drop in the recall: a decrease of 22.98\% on \autoS and 32.21\% on \realS. Furthermore, RL training enhances recall by 6.24\% on \autoS and 19.96\% on \realS. The RL training curves are depicted in Figure~\ref{score_and_loss} in Appendix~\ref{ppo_training_detail}, where the training curves show a steady increase in return with the training steps, eventually converging after 200 steps. Finally, removing the Selector as an auxiliary reward model results in a 3.76\% recall drop on \autoS and a 9.63\% drop on \realS.

We investigate how to control agent behavior by adjusting the rewards in RL training. Experiments are conducted with varying reward coefficients $\alpha$ in Equation~\ref{action_reward}, and results are presented in Table~\ref{ablation_reward}. We report two metrics: crawler recall and crawler action. The crawler action refers to the total number of \texttt{[Search]} and \texttt{[Expand]} actions throughout the Crawler’s entire trajectory. As the reward increases, both crawler recall and crawler action increase, suggesting that adjusting rewards in RL training can effectively influence \pasa's behavior.

\begin{table}[htbp]
\renewcommand{\arraystretch}{1}
\centering
\scalebox{0.9}{
\begin{tabular}{ccccc}
\toprule[1pt]
\multirow{2}{*}{\textbf{$\alpha$}} & \textbf{Crawler} & \textbf{Crawler} 
 & \multirow{2}{*}{\textbf{Precision}} & \multirow{2}{*}{\textbf{Recall}}\\ & \textbf{Recall} & \textbf{Actions} & & \\
\midrule[0.5pt]
0.5 & 0.7227 & 175.9 & 0.1458 & 0.4602 \\
1.0 & 0.7708 & 319.8 & 0.1451 & 0.4792\\
1.5 & 0.7931 & 382.4 & 0.1448 & 0.4834\\
2.0 & 0.8063 & 785.5 & 0.1409 & 0.4869\\
\bottomrule[1pt]
\end{tabular}
}
\caption{Performance of the Crawler trained on different reward coefficient $\alpha$ on AutoScholarQuery test set.}
\label{ablation_reward}
\vspace{-0.2cm}
\end{table}

\section{Conclusion}

In this paper, we introduce \pasa, a novel paper search agent designed to provide comprehensive and accurate results for complex academic queries. \pasa is implemented within the AGILE, a reinforcement learning framework for LLM agents. To train \pasa, we developed \autoS, a dataset of fine-grained academic queries and corresponding papers drawn from top-tier AI conference publications. To evaluate \pasa in real-world scenarios, we also constructed \realS, a dataset of actual academic queries paired with annotated papers. Our experimental results demonstrate that \pasa outperforms all baselines, including Google, Google Scholar, and Google with GPT-4o, ChatGPT, GPT-o1, and \pasa-GPT-4o. In particular, \pasa-7B surpasses Google with GPT-4o by 37.78\% in recall@20 and 39.90\% in recall@50, while also exceeding \pasa-GPT-4o by 30.36\% in recall and 4.25\% in precision. These findings underscore that \pasa significantly improves the efficiency and accuracy of academic search. 

\section*{Limitations}

(1) Our dataset collection and experiments were primarily focused on the field of machine learning. Although our proposed method is general, we did not explore its performance in other scientific fields. We leave to investigate its applicability to other domains in future work.

\noindent (2) Due to resource constraints, our experiments primarily use LLMs with 7b parameters. We expect that scaling up to larger models will lead to more powerful agents. Expanding PaSa to leverage larger LLMs is our future work.

\section*{Acknowledgments}
The authors thank Yaohua Fang, Zheng Li, Qiang Lu, Yelong Shi, Xuguang Wei, and Tingshuai Yan from ByteDance for their support in developing the PaSa demo. We also thank Jianghui Xie at ByteDance for her assistance with the release of the PaSa demo. Finally, we thank the anonymous reviewers for their valuable suggestions that helped improve this work.

\bibliography{custom}

\appendix

\begin{table*}[!h]
\centering
\scalebox{0.8}{
\begin{tabular}{|p{1.25\textwidth}|}
\hline
\begin{tabular}[c]{@{}p{1.25\textwidth}@{}} \textbf{Query:} Give me papers about how to rank search results by the use of LLM\\ 
\textbf{Query Date:} 2024-10-01\\ 
\textbf{Answer Papers:} 
\\ {[}0{]} Instruction Distillation Makes Large Language Models Efficient Zero-shot   Rankers (2311.01555)
\\ {[}1{]} Beyond Yes and No: Improving Zero-Shot LLM Rankers via Scoring   Fine-Grained Relevance Labels (2310.14122)
\\ {[}2{]} Large Language Models are Effective Text Rankers with Pairwise Ranking   Prompting (2306.17563)
\\ {[}3{]} A Setwise Approach for Effective and Highly Efficient Zero-shot Ranking   with Large Language Models (2310.09497)
\\ {[}4{]} RankVicuna: Zero-Shot Listwise Document Reranking with Open-Source Large   Language Models (2309.15088)
\\ {[}5{]} PaRaDe: Passage Ranking using Demonstrations with Large Language Models (2310.14408)
\\ {[}6{]} Is ChatGPT Good at Search? Investigating Large Language Models as   Re-Ranking Agents (2304.09542)
\\ {[}7{]} Large Language Models are Zero-Shot Rankers for Recommender Systems (2305.08845)
\\ {[}8{]} TourRank: Utilizing Large Language Models for Documents Ranking with a Tournament-Inspired Strategy (2406.11678)
\\ {[}9{]} ExaRanker: Explanation-Augmented Neural Ranker (2301.10521)
\\ {[}10{]} RankRAG: Unifying Context Ranking with Retrieval-Augmented Generation in LLMs (2407.02485)
\\ {[}11{]} Make Large Language Model a Better Ranker (2403.19181)
\\ {[}12{]} LLM-RankFusion: Mitigating Intrinsic Inconsistency in LLM-based Ranking (2406.00231)
\\ {[}13{]} Improving Zero-shot LLM Re-Ranker with Risk Minimization (2406.13331)
\\ {[}14{]} Zero-Shot Listwise Document Reranking with a Large Language Model (2305.02156)
\\ {[}15{]} Consolidating Ranking and Relevance Predictions of Large Language Models through Post-Processing (2404.11791)
\\ {[}16{]} Re-Ranking Step by Step: Investigating Pre-Filtering for Re-Ranking with Large Language Models (2406.18740)
\\ {[}17{]} Large Language Models for Relevance Judgment in Product Search (2406.00247)
\\ {[}18{]} PromptReps: Prompting Large Language Models to Generate Dense and Sparse Representations for Zero-Shot Document Retrieval (2404.18424)
\\ {[}19{]} Passage-specific Prompt Tuning for Passage Reranking in Question Answering with Large Language Models (2405.20654)
\\ {[}20{]} When Search Engine Services meet Large Language Models: Visions and Challenges (2407.00128)
\\ {[}21{]} RankZephyr: Effective and Robust Zero-Shot Listwise Reranking is a   Breeze! (2312.02724)
\\ {[}22{]} Rank-without-GPT: Building GPT-Independent Listwise Rerankers on   Open-Source Large Language Models (2312.02969)
\\ {[}23{]} MuGI: Enhancing Information Retrieval through Multi-Text Generation Integration with Large Language Models (2401.06311)
\\ {[}24{]} Discrete Prompt Optimization via Constrained Generation for Zero-shot   Re-ranker (2305.13729)
\\ {[}25{]} REAR: A Relevance-Aware Retrieval-Augmented Framework for Open-Domain   Question Answering (2402.17497)
\\ {[}26{]} Agent4Ranking: Semantic Robust Ranking via Personalized Query Rewriting   Using Multi-agent LLM (2312.15450)
\\ {[}27{]} FIRST: Faster Improved Listwise Reranking with Single Token Decoding (2406.15657)
\\ {[}28{]} Leveraging LLMs for Unsupervised Dense Retriever Ranking (2402.04853)
\\ {[}29{]} Unsupervised Contrast-Consistent Ranking with Language Models (2309.06991)
\\ {[}30{]} Enhancing Legal Document Retrieval: A Multi-Phase Approach with Large Language Models (2403.18093)
\\ {[}31{]} Found in the Middle: Permutation Self-Consistency Improves Listwise Ranking in Large Language Models (2310.07712)
\\ {[}32{]} Fine-Tuning LLaMA for Multi-Stage Text Retrieval (2310.08319)
\\ {[}33{]} Zero-shot Audio Topic Reranking using Large Language Models (2309.07606)
\\ {[}34{]} Uncovering ChatGPT's Capabilities in Recommender Systems (2305.02182)
\\ {[}35{]} Cognitive Personalized Search Integrating Large Language Models with an Efficient Memory Mechanism (2402.10548)
\\ {[}36{]} Towards More Relevant Product Search Ranking Via Large Language Models: An Empirical Study (2409.17460)
\\ {[}37{]} Pretrained Language Model based Web Search Ranking: From Relevance to Satisfaction (2306.01599)
\\ {[}38{]} Open-source large language models are strong zero-shot query likelihood models for document ranking (2310.13243)\end{tabular} \\ \hline
\end{tabular}
}
\caption{Examples of queries and corresponding papers in \realS.}
\label{tab:realS_examples}
\end{table*}

\section{Quality Evaluation of \autoS}\label{auto_eval}








To assess the quality of \autoS, we sampled 100 query-paper pairs and evaluated the rationality and relevance of each query and its corresponding paper. The detailed evaluation criteria are as follows:

\begin{itemize}
\item A qualified query should be a complete and understandable sentence. For example, incomplete or fragmented sentences are not acceptable.

\item A query that misrepresents the meaning of the source paper, leading to irrelevant citations, is not qualified. This includes queries that exaggerate the scope or introduce incorrect conditions.

\item A query is ambiguous if there is insufficient context and additional information is needed. For instance, abbreviations with multiple meanings can create ambiguity, leading to the corresponding citations being incomplete answer lists.

\item An answer paper is considered qualified if it aligns with the requirements of the query. The paper should address all or most of the essential factors that make it a suitable response. 
\end{itemize}

Our quality check found that 94.0\% of the queries were qualified. Among them, 93.7\% of the corresponding answer papers were also qualified. The primary reason for unqualified papers was inaccurate citations in the source paper.

\begin{table*}[htbp]
\renewcommand{\arraystretch}{1}
\centering
\resizebox{\linewidth}{!}{%
\begin{tabular}{p{19cm}}
\toprule[1pt]
\textbf{The prompt for search query generation}\\
\midrule[0.5pt]
You are an elite researcher in the field of AI, please generate some mutually exclusive queries in a list to search the relevant papers according to the User Query. Searching for a survey paper would be better.

User Query: \{user\_query\}

The semantics between generated queries are not mutually inclusive. The format of the list is: [``query1'', ``query2'', ...]

Queries:\\
\bottomrule[1pt]
\end{tabular}
}
\caption{The prompt for GPT-4o to generate search queries from the user query.}
\label{gen_search_query}
\end{table*}

\begin{table*}[htbp]
\renewcommand{\arraystretch}{1}
\centering
\resizebox{\linewidth}{!}{%
\begin{tabular}{p{1.4cm}p{7.6cm}p{10cm}}
\toprule[1pt]
&\textbf{Search Session starting from $S_q$}&\textbf{Expand Session starting from $S_{q+p}$}\\
\midrule[0.5pt]
prompt&Please, generate some mutually exclusive queries in a list to search the relevant papers according to the User Query. Searching for survey papers would be better.

User Query: \{user\_query\}&
You are conducting research on \textquotesingle\{user\_query\}\textquotesingle. You need to predict which sections to look at to get more relevant papers. 

Title: \{title\}

Abstract: \{abstract\}

Sections: \{sections\}\\
\midrule[0.5pt]

response&[Search] \{query 1\}

[Search] \{query 2\}

...

[Stop]&
[Expand] \{section 1\}

[Expand] \{section 2\}

...

[Stop]\\
\bottomrule[1pt]
\end{tabular}
}
\caption{The session trajectory templates of the Crawler.}
\label{crawler_prompt}
\end{table*}

\section{Annotation details}\label{annotation_details}

The annotators of \realS include professors from the Department of Computer Science at a top-tier university in China. They are paid \$4 per data entry, which consists of a user query and a research paper. Their task is to determine whether the paper satisfies the query. 

\subsection{Annotation Instructions}
For each <user query, paper> pair, carefully assess whether the paper address the user query. Write your decision and provide a brief explanation (1-2 sentences). Specific guidelines are as follows:
\begin{itemize}
    \item You may read the entire paper to determine whether it satisfies the user query.
    \item Exclude survey papers unless the user query specifically requests them.
    \item All conditions in the user query must be met for the paper to be considered qualified.
\end{itemize}

\subsection{Quality control}

The annotation process follows the following quality control measures:

\begin{itemize}
    \item Stage 1: Annotators work in batches of 20. Authors review 100\% of the annotations. Once the consistency rate on the first pass reaches 90\%, the process moves to Stage 2.

    \item Stage 2: Annotators work in batches of 50. Authors randomly check 40\% of the annotations. If the consistency rate is below 90\%, the entire batch is re-annotated and re-checked. Once the batch meets the 90\% consistency rate on the first pass, the process moves to Stage 3.

    \item Stage 3: Annotators work in batches of 100. Authors randomly check 20\% of the annotations. If the consistency rate is below 90\%, the entire batch is re-annotated and re-checked.
\end{itemize}

Two authors conducted the quality control.

\section{Example in \realS}\label{data_example}

Table~\ref{tab:realS_examples} presents an example query and corresponding papers from \realS.

\section{Implementation Details of the Crawler}

\subsection{Imitation learning data generation}\label{crawler_detail}

We generate training data for imitation learning on a session-by-session basis. There are two types of sessions: \emph{search session} (starting from state $S_q$) and \emph{expand session} (starting from state $S_{q+p}$). 

For search sessions starting from $S_q$, we sample user queries from the \autoS training set and prompt GPT-4o to generate corresponding search queries. The prompt template is shown in Table~\ref{gen_search_query}. The session trajectory is constructed by adding a \texttt{[Search]} token before each query, concatenating the queries, and appending a \texttt{[Stop]} token at the end, as shown in Table~\ref{crawler_prompt}. A total of 3,011 search session trajectories are generated.

For expanded sessions starting from $S_{q+p}$, we continue by searching for the generated queries using Google. We then sample papers from the search results and obtain the initial state, which includes both the query and a paper. To build the session trajectory, we examine each sub-section of the paper. If the sub-section references at least one paper in the \autoS training set corresponding to the query, the sub-section is selected. Otherwise, the sub-section is selected with a 10\% probability to enhance trajectory diversity. The selected sections are filled into the template in Table~\ref{crawler_prompt}, completing the session trajectory. In total, 9,978 expanded session trajectories are constructed.

\subsection{PPO training}\label{ppo_training_detail}
During PPO training, each device processes 4 user queries in each step, generating a search session for each user query. Then, 6 expansion sessions are created by randomly sampling 6 papers from the search results. This process is repeated with the expanded citation results, yielding 6 additional expanded sessions. In total, 16 session trajectories are generated per step.

\begin{table}[htbp]
\renewcommand{\arraystretch}{1}
\centering
\scalebox{0.95}{
\begin{tabular}{lll}
\toprule[1pt]
\multicolumn{2}{c}{\textbf{Name}} & \textbf{Value} \\
\midrule[0.5pt]
$\alpha$ & (Equation~\ref{action_reward}) & 1.5 \\
$c(\text{[Search]})$ & (Equation~\ref{action_reward}) & 0.1 \\
$c(\text{[Expand]})$ & (Equation~\ref{action_reward}) & 0.1 \\
$c(\text{[Stop]})$ & (Equation~\ref{action_reward}) & 0.0 \\
$\gamma_0$ & (Equation~\ref{q_function}) & 1.0 \\
$\gamma_1$ & (Equation~\ref{q_function}) & 0.1 \\
$\beta$ & (Equation~\ref{q_function}) & 0.1 \\
$\epsilon$ & (Equation~\ref{policy_loss},~Equation~\ref{value_loss}) & 0.2 \\
$\eta$ & (Equation~\ref{rl_loss}) & 10 \\
\multicolumn{2}{l}{learning rate} & 1e-6 \\
\multicolumn{2}{l}{epoch per step} & 2 \\
\multicolumn{2}{l}{forward batch size} & 1 \\
\multicolumn{2}{l}{accumulate batch size} & 16 \\
\multicolumn{2}{l}{NVIDIA H100 GPU} & 16 \\
\multicolumn{2}{l}{policy freezing step} & 50 \\
\multicolumn{2}{l}{total step} & 250 \\
\bottomrule[1pt]
\end{tabular}
}
\caption{The hyperparameters used in PPO training.}
\label{tab:crawler_hyperparameters}
\end{table}

\begin{figure}[h]
    \centering
    \includegraphics[width=1.0\linewidth]{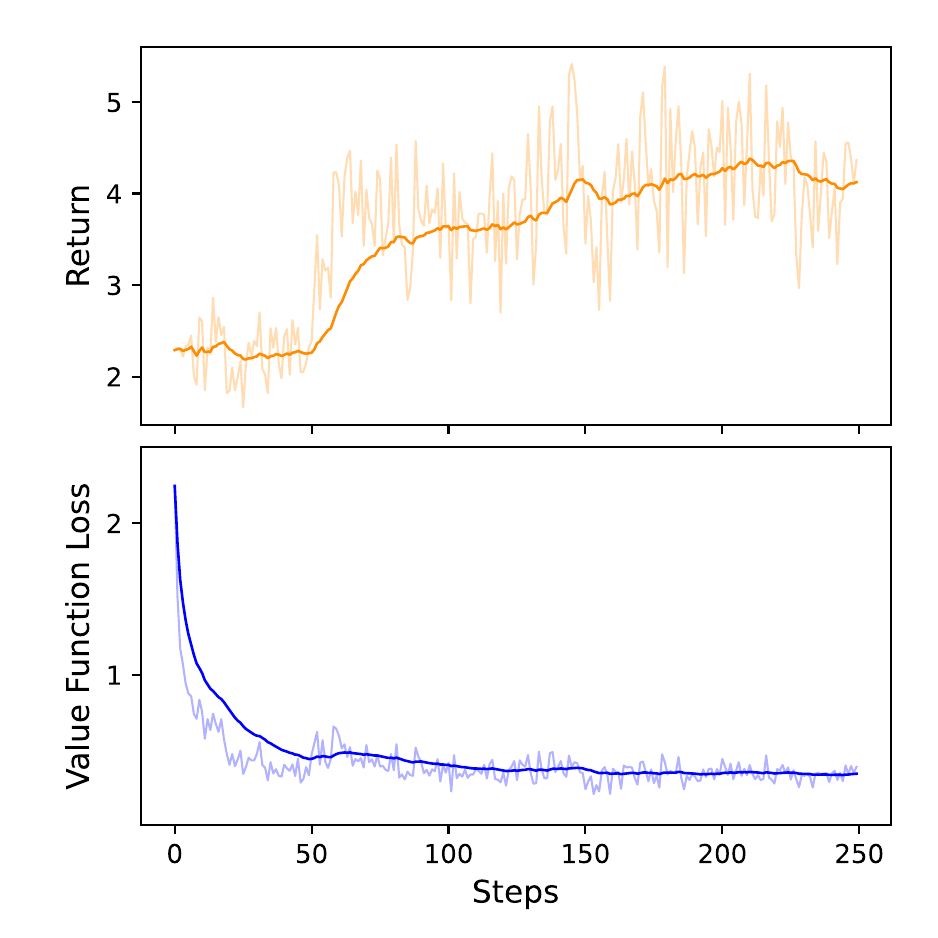}
    \caption{Return and value function loss curves during the PPO training process. The smoothing method of the curve in the figures is the exponential moving average(EMA) formula that aligns with the one used in TensorBoard, and the smoothing weight is set to 0.95.}
    
    \label{score_and_loss}
\vspace{-8pt}
\end{figure}

Table~\ref{tab:crawler_hyperparameters} lists the hyperparameters used during the training process. 
Figure~\ref{score_and_loss} depicts the RL training curves, which show a steady increase in return with the training steps, eventually converging after 200 steps.

\section{Implementation Details of the Selector}\label{selector_detail}

\begin{table*}[h]
\renewcommand{\arraystretch}{1}
\centering
\resizebox{\linewidth}{!}{%
\begin{tabular}{p{19cm}}
\toprule[1pt]
\textbf{The prompt for paper selection}\\
\midrule[0.5pt]
You are an elite researcher in the field of AI, conducting research on \{user\_query\}. Evaluate whether the following paper fully satisfies the detailed requirements of the user query and provide your reasoning. Ensure that your decision and reasoning are consistent.

Searched Paper:

Title: \{title\}

Abstract: \{abstract\}

User Query: \{user\_query\}

Output format: Decision: True/False

Reason:... 

Decision:\\
\bottomrule[1pt]
\end{tabular}
}
\caption{Prompt used by \pasa Selector or GPT-4o to evaluate paper relevance.}
\label{tab:selector_prompt}
\end{table*}

\begin{table*}[htbp]
\renewcommand{\arraystretch}{1}
\centering
\scalebox{0.85}{
\begin{tabular}{lcccccc}
\toprule[1pt]
\multicolumn{1}{l}{\textbf{Method}} & \textbf{Crawler Recall} & \textbf{Precision} & \textbf{Recall} & \textbf{Recall@100} & \textbf{Recall@50} & \textbf{Recall@20} \\
\midrule[0.5pt]
Google & - & - & - & 0.2101 & 0.2010 & 0.1788\\
Google Scholar & - & - & - & 0.0801 & 0.0739 & 0.0612\\
Google with GPT-4o & - & - & - & 0.2101 & 0.2010 & 0.1788\\
ChatGPT & - & 0.0507 & 0.3046 & - & - & - \\
GPT-o1 & - & 0.0374 & 0.2006 & - & - & - \\
\pasa-GPT-4o & 0.7595 & 0.1817 & 0.4488 & - & - & - \\
\midrule[0.5pt]
\pasa-7b & \textbf{0.7752} & \textbf{0.1881} & \textbf{0.5328} & \textbf{0.6932} & \textbf{0.6543} & \textbf{0.5494} \\
\pasa-7b-ensemble & \textbf{0.8244} & \textbf{0.1822} & \textbf{0.5568} & \textbf{0.7041} & \textbf{0.6795} & \textbf{0.5535} \\
\bottomrule[1pt]
\end{tabular}
}
\caption{Results on 100-sample subset of \autoS test.}
\label{main_results_100}
\vspace{-0.2cm}
\end{table*}

We begin by sampling user queries from the \autoS training set. For each user query and one of its corresponding papers in the \autoS training set, we prompt GPT-4o to generate a decision token and rationale (see Table~\ref{tab:selector_prompt} for the prompt). We reject any data where the decision token is "False", as this contradicts the \autoS label. The remaining data are retained as positive <user query, paper> pairs.

Next, we simulate a partial paper search using \pasa-GPT-4o. In this simulation, each paper has a 50\% probability of being added to the paper queue. Pairs where the paper is not selected by GPT-4o and is not in the \autoS training set are labeled as negative examples.

The final training dataset consists of 19,812 <user query, paper> pairs, each with a decision token and rationale generated by GPT-4o, drawn from 9,000 instances in the \autoS training set.

\section{Selector Test Dataset} \label{sec:Details of Selector Evaluation}

We select 200 queries from the \autoS development set. For each query, we perform a Google search and randomly choose one paper from the union of the search results and the relevant paper set in \autoS. This yields a set of <user query, paper> pairs. Annotators then evaluate whether each paper aligns with the requirements of the user query. The final test dataset consists of 98 positive samples and 102 negative samples.



\begin{table}[htbp]
\renewcommand{\arraystretch}{1}
\centering
\scalebox{0.9}{
\begin{tabular}{ccccc}
\toprule[1pt]
\multirow{2}{*}{\textbf{$c(a_t)$}} & \textbf{Crawler} & \textbf{Crawler} 
 & \multirow{2}{*}{\textbf{Precision}} & \multirow{2}{*}{\textbf{Recall}}\\ & \textbf{Recall} & \textbf{Actions} & & \\
\midrule[0.5pt]
0 & 0.8239 & 1296.3 & 0.1388 & 0.4852 \\
0.1 & 0.7931 & 382.4 & 0.1448 & 0.4834\\
0.2 & 0.7478 & 230.1 & 0.1489 & 0.4764\\
\bottomrule[1pt]
\end{tabular}
}
\caption{Performance of the Crawler trained on different action cost $c(a_t)$ on AutoScholarQuery test set.}
\label{ablation_cost}
\vspace{-0.2cm}
\end{table}

\section{Additional Experimental Results}\label{exp_rst_appx}

\subsection{Results on 100-sample subset of \autoS}

To ensure a fair comparison with the ChatGPT baseline, which is evaluated on only 100 samples from \autoS test, we report the performance of all methods on the same subset in Table~\ref{main_results_100}. The results align with those in Table~\ref{main_results}, confirming that \pasa-7b consistently outperforms all baselines.


\subsection{Action cost}

We incorporate action costs to prevent the agent from taking excessive, unproductive actions. Without such costs, the total number of actions would increase significantly without yielding meaningful outcomes. 

The key consideration is the reward coefficient $\alpha$ and the action cost $c(a_t)$ in Equation~\ref{action_reward}. In Table~\ref{ablation_reward}, we fix $c(a_t)$ and analyze how varying $\alpha$ affects performance.

Additionally, Table~\ref{ablation_cost} shows how different values of $c(a_t)$ affect the final performance.

\section{Prompt Templates}

\subsection{Prompts used for data synthesis in \autoS}\label{autoS_prompt}

Table~\ref{tab:AutoScholarQuery_prompt} presents the prompt template used with GPT-4o to automatically generate \autoS. For each paper, we extract its \emph{Related Work} section, input it into GPT-4o, and use the prompt to generate scholarly queries along with their corresponding paper answers.

\subsection{Prompts for baselines}\label{prompt_templates}

Table~\ref{tab:search_query_paraphrase_prompt} presents the search query paraphrasing prompt used for the baseline Google with GPT-4o.

Table~\ref{tab:prompt_Chatgpt}, \ref{tab:prompt_o1} and \ref{prompt_pasa_gpt4o} show the prompts used for the ChatGPT baseline (search-enabled GPT-4o), the GPT-o1 baseline and \pasa-GPT-4o, respectively.


\begin{table*}[h]
\renewcommand{\arraystretch}{1}
\centering
\resizebox{\linewidth}{!}{%
\begin{tabular}{p{19cm}}
\toprule[1pt]
\textbf{The prompt for \autoS generation}\\
\midrule[0.5pt]
You are provided a `Related Work' section of a research paper. The researcher reviewed the relevant work, conducted a literature survey, and cited corresponding references in this text (enclosed by `\textbackslash cite' tags with IDs). Can you guess what research questions the researcher might have posed when preparing this text? The answers to these questions should be the references cited in this passage. Please list questions and provide the corresponding answers.

[Requirements:]

1. Craft questions similar to those a researcher would pose when reviewing related works, such as ``Which paper studied ...?'', ``Any works about...?'', ``Could you provide me some works...?''

2. Construct the question-answer pairs based on [Section from A Research Paper]. The answer should be the cited papers in [Section from A Research Paper]. 

3. Do not ask questions including "or" or "and" that may involve more than one condition.

4. Clarity: Formulate questions clearly and unambiguously to prevent confusion.

5. Contextual Definitions: Include explanations or definitions for specialized terms and concepts used in the questions.

6. Format the output as a JSON array containing five objects corresponding to the three question-answer pairs. 

Here are some examples:

[Begin of examples]

\{Section from A Research Paper-1\}

\{OUTPUT-1\}

\{Section from A Research Paper-2\}

\{OUTPUT-2\}

\{Section from A Research Paper-3\}

\{OUTPUT-3\}

[End of examples]

\{Section from A Research Paper\}

[OUTPUT]: \\
\bottomrule[1pt]
\end{tabular}
}
\caption{The prompt used with GPT-4o to automatically synthesize \autoS.}
\label{tab:AutoScholarQuery_prompt}
\end{table*}

\begin{table*}[h]
\renewcommand{\arraystretch}{1}
\centering
\resizebox{\linewidth}{!}{%
\begin{tabular}{p{19cm}}
\toprule[1pt]
\textbf{The prompt for search query paraphrase}\\
\midrule[0.5pt]
Generate a search query suitable for Google based on the given academic paper-related query. Here’s the structure and requirements for generating the search query:

Understand the Query: Read and understand the given specific academic query.

Identify Key Elements: Extract the main research field and the specific approaches or topics mentioned in the query.

Formulate the Search Query: Combine these elements into a concise query that includes terms indicating academic sources.
Do not add any site limitations to your query.

[User's Query]: \{user\_query\}

[Generated Search Query]: \\
\bottomrule[1pt]
\end{tabular}
}
\caption{The search query paraphrasing prompt used for the Google with GPT-4o baseline.}
\label{tab:search_query_paraphrase_prompt}
\end{table*}

\begin{table*}[h]
\renewcommand{\arraystretch}{1}
\centering
\resizebox{\linewidth}{!}{%
\begin{tabular}{p{19cm}}
\toprule[1pt]
\textbf{The prompt for ChatGPT (search-enabled GPT-4o)}\\
\midrule[0.5pt]
[User's Query]

You should return the Arxiv papers. You should provide more than 10 papers you searched  in JSON format: 

\{"paper\_1": \{"title": , 'authors': , 'link': \}, "paper\_2": \{"title": , 'authors': , 'link': \}\}
 \\
\bottomrule[1pt]
\end{tabular}
}
\caption{The prompt for ChatGPT baseline (search-enabled GPT-4o).}
\label{tab:prompt_Chatgpt}
\end{table*}

\begin{table*}[h]
\renewcommand{\arraystretch}{1}
\centering
\resizebox{\linewidth}{!}{%
\begin{tabular}{p{19cm}}
\toprule[1pt]
\textbf{The prompt for GPT-o1}\\
\midrule[0.5pt]
\{user\_query\}

You should return arxiv papers. You should provide more than 10 paper you searched in JSON format: \{"paper\_1": \{"title": , "authors": , "link": \}, "paper\_2": \{"title": , "authors": , "link": \}\}. Do not return any other description.
 \\
\bottomrule[1pt]
\end{tabular}
}
\caption{The prompt for GPT-o1 baseline.}
\label{tab:prompt_o1}
\end{table*}

\begin{table*}[h]
\renewcommand{\arraystretch}{1}
\centering
\resizebox{\linewidth}{!}{%
\begin{tabular}{p{19cm}}
\toprule[1pt]
\textbf{The prompt for search session of Crawler}\\
\midrule[0.5pt]
You are an elite researcher in the field of AI, please generate some mutually exclusive queries in a list to search the relevant papers according to the User Query. Searching for a survey paper would be better.

User Query: \{user\_query\}

The semantics between generated queries are not mutually inclusive. The format of the list is: [``query1'', ``query2'', ...]

Queries:\\
\midrule[0.5pt]
\textbf{The prompt for the expand session of Crawler}\\
\midrule[0.5pt]
You are an elite researcher in the field of AI, currently conducting research on the [topic] specified below. Your task is to determine if the paper specified below likely contains highly relevant citations for the [topic] and, if so, to identify specific sections where these citations might appear.

Task Instructions:

1. Relevance Assessment: Decide if the paper is likely to include citations highly relevant to the given [topic]. Output "Yes" or "No" on the first line.

2. Section Selection: If you answered "Yes" in step 1, identify which sections of the paper are likely to contain these relevant citations. From the list of provided sections, select only those you think may contain relevant citations. If no sections seem relevant even if your answer to step 1 was "Yes," leave this empty. Output the selected sections in JSON format on the second line.

[topic]: \{user\_query\}

[paper title]: \{title\}

[paper abstract]: \{abstract\}

[paper sections]: \{sections\}

Output Format: Output exactly two lines:

1. The first line: Either "Yes" or "No" based on the relevance assessment.

2. The second line: A JSON string with selected sections, e.g., \{\{"selected\_section\_1": section\_name\_1, "selected\_section\_2": section\_name\_2\}\}. If no sections are selected, output \{\{\}\}.\\
\midrule[0.5pt]
\textbf{The prompt for Selector}\\
\midrule[0.5pt]
You are an elite researcher in the field of AI, conducting research on \{user\_query\}. Evaluate whether the following paper fully satisfies the detailed requirements of the user query and provide your reasoning. Ensure that your decision and reasoning are consistent.

Searched Paper:

Title: \{title\}

Abstract: \{abstract\}

User Query: \{user\_query\}

Output format: Decision: True/False

Reason:... 

Decision:\\
\bottomrule[1pt]
\end{tabular}
}
\caption{The prompts for \pasa-GPT-4o.}
\label{prompt_pasa_gpt4o}
\end{table*}


\end{document}